\documentclass{emulateapj}
\usepackage{graphics}

\def\msun{\,  {\rm M_\odot}}
\def\cm-3{\,{\rm cm^{-3}}}

\def\kpc-3{\,{\rm kpc^{-3}}}
\def\myr-1{\,{\rm Myr^{-1}}}

\def\kpc{\,{\rm kpc}}

\def\siggas{$\Sigma_{\rm{gas}}\ $}

\def\NOVI{$N_{\rm{OVI}}\ $}
\def\sigSFR{$\dot{\Sigma}_{\rm{SFR}}$}

\usepackage{natbib}
\usepackage{color}
\usepackage{rotating}
\usepackage{footnote}
\usepackage{datetime}
\usepackage{amsmath}
\usepackage{mathrsfs}
\usepackage[normalem]{ulem}
\begin{document}

\title{OVI emission from the supernovae-regulated interstellar medium: \\ Simulation vs Observation  }

\author{Miao Li\altaffilmark{$\dagger$,1}, Greg L. Bryan\altaffilmark{1,2}, Jeremiah P. Ostriker\altaffilmark{1} } 

\altaffiltext{$\dagger$}{miao@astro.columbia.edu}
\altaffiltext{1}{Department of Astronomy, Columbia University, 550 W120th Street, New York, NY 10027, USA} 
\altaffiltext{2}{Simons Center for Computational Astrophysics, 160 5th Ave, New York, NY, 10010, USA}

\begin{abstract}
{
The OVI $\lambda\lambda$1032, 1038\AA\ doublet emission traces collisionally ionized gas with $T\approx 10^{5.5}$ K, where the cooling curve peaks for metal-enriched plasma. This warm-hot phase is usually not well-resolved in numerical simulations of the multiphase interstellar medium (ISM), but can be responsible for a significant fraction of the emitted energy.  Comparing simulated OVI emission to observations is therefore a valuable test of whether simulations predict reasonable cooling rates from this phase. We calculate OVI $\lambda$1032\AA\ emission, assuming collisional ionization equilibrium, for our small-box simulations of the stratified ISM regulated by supernovae.  We find that the agreement is very good for our solar neighborhood model, both in terms of emission flux and mean OVI density seen in absorption.  We explore runs with higher surface densities and find that, in our simulations, the OVI emission from the disk scales roughly linearly with the star formation rate.  Observations of OVI emission are rare for external galaxies, but our results do not show obvious inconsistency with the existing data.  Assuming  the solar metallicity, OVI emission from the galaxy disk in our simulations accounts for roughly 0.5\% of supernovae heating.

 }

\end{abstract}

\section{Introduction}
\label{intro}

Supernova (SN) explosions represent a key form of stellar feedback in galaxies, and produce copious hot gas in the ISM. The SN energy can regulate the ISM dynamics and launch galactic outflows \citep{cox74, mckee77, cox05}. One key factor to quantify the efficiency of SN feedback is the energy loading factor (or ``thermalization factor''), which is the fraction of SN energy that is loaded into the outflows. The efficiency of radiative cooling in the multiphase ISM directly affects this loading efficiency. 

A temperature around $10^{5.5}$ K is where the cooling curve peaks for metal-enriched gas \citep[e.g.][]{sutherland93}. The OVI $\lambda\lambda$1032, 1038\AA\ doublet is one major coolant around this temperature.  These OVI resonant lines have been observed widely, from the local ISM, which provides the first strong evidence of the prevalence of the hot phase in the ISM \citep{jenkins74, york74}, to the circumgalactic medium (CGM) and even the intergalactic medium (IGM) \citep[e.g.][]{ bregman06,tumlinson11, simcoe02}.   Most of these observations, though, focus on absorption instead of emission, except for the solar neighborhood \citep[e.g.][]{otte03}. The origin of gas around this temperature is not totally clear. Thermal conduction/turbulent mixing at the interface between the hot plasma ($T \gtrsim 10^6$ K) and warm clouds ($T \sim 10^4$ K) \citep{mckee77,weaver77,cowie79, begelman90}, and cooling from hotter gas \citep{edgar86, slavin93, dopita96, heckman02} have been suggested as possible mechanisms. 

In hydrodynamic simulations of astrophysical systems, the quantity of gas around $10^{5.5}$ K can be affected by numerical mixing and diffusion. Depending on the resolution and technique used, hydrodynamic simulations can artificially promote or suppress the mixing and the amount of gas in the warm-hot phase. For example, the particle-based SPH scheme usually exhibits too little mixing, while the mesh-based techniques can have too much mixing \citep{agertz07,springel10}. The magnitude of this numerical artifact is hard to gauge. This is a particular issue when multiple phases -- with contrasting densities and temperatures -- coexist at similar pressures, as is the case for the ISM. Even without explicitly including thermal conduction, gas with intermediate temperature $10^5-10^6$ K can show up around the interface between the hot and warm (or cold) phases, which is usually not well-resolved.  Although higher numerical resolution helps, turbulence may drive the scale of the cold/hot interface down to small length scales and, until the Field length is resolved, improved resolution may simply result in more clouds without reducing the amount of material in the interface.  Because of the high cooling efficiency in this warm-hot phase, it is important to test the simulation cooling rates.  One way to do this is to compare against observed emission lines, which trace the cooling directly.  We make such an attempt in this Letter. 

The OVI $\lambda\lambda$1032,1038\AA\ doublet falls in the far-UV band, which cannot be observed from the ground. Satellites with FUV spectrographs, such as \textit{Corpernicus, FUSE, SPEAR} and \textit{HST}, contribute to most of our knowledge about gas traced by OVI. For the local ISM, the column density of OVI is obtained from absorption-line studies of bright stars \citep{jenkins78a, jenkins78b, oegerle05, savage06}. OVI line emission has also been detected in various sight lines \citep{dixon06,otte06, welsh07}. 

For external galaxies, the OVI interstellar absorption line is seen in nearly every starburst system observed \citep [e.g.][]{grimes07}. The CGM around star forming galaxies shows absorption of OVI out to $\gtrsim 100$ kpc \citep{tumlinson11}. OVI emission from the disk or the CGM, on the other hand, has only been reported in a few cases \citep{otte03, grimes07, hayes16}.

In this Letter, we calculate the emission of OVI 1032\AA\, as well as the OVI column densities, for our ISM simulations with SN feedback. \cite{deavillez05, deavillez12} simulated the SNe-regulated ISM for the solar neighborhood, and found OVI absorption agreeing well with the observations. We extend that study by calculating both the emission and the column density, for conditions with various star formation rates.  We describe the simulations and calculation of OVI emission in Section 2, show the results and compare with the observations in 3, and summarize in 4. 

\section{Methods}
\label{sec:method}

The simulations presented in this paper are the four fiducial runs in  \cite{li16}, hereafter referred to as Paper I. We briefly introduce the setup here -- for details, see Paper I.

Our simulations are performed using the Eulerian hydrodynamic code Enzo \citep{bryan14}. We use the higher-order piece-wise parabolic method \citep[PPM][]{colella84} as the hydro-solver, along with the two-shock Riemann solver.
The 3D, rectangular simulation boxes cover a fraction of the galaxy disk, with a square cross-section and the long z-axis perpendicular to the disk plane. Similar configurations have been widely adopted \citep[e.g.][]{dib06, joung06, creasey13, walch15, kim16}. The four runs discussed in this paper have initial gas surface densities \siggas $=$ 1, 10, 55, 150 $\msun\ \rm{pc}^{-2}$, respectively, referred to as $\Sigma$1-KS, $\Sigma$10-KS, $\Sigma$55-KS, $\Sigma$150-KS. The simulation $\Sigma$10-KS is the model for the solar neighborhood. A static gravitational field from gas, stars and the dark matter halo is included.

The midplane is located at $z=0$. The box size in the z-direction is 5 kpc, i.e., --2.5 $\leqslant z \leqslant$ 2.5 kpc. The spatial resolution at $|z|\leqslant$ 0.5 kpc is 5.0, 2.0, 0.75, and 0.60 pc, respectively, so simulations with larger \siggas have higher resolutions. The resolution is coarsened by a factor of 2 for $0.5\leqslant |z| \leqslant$ 1 kpc, and by another factor of 2 for $|z|\geqslant$ 1 kpc. An outflowing boundary condition is applied for the z-direction, and periodic boundaries for the x- and y-directions. 

We have chosen the star formation surface density \sigSFR\ to be $1.26\times 10^{-4}$, $6.31\times 10^{-3}$, 0.158, 0.708 $\msun\ \rm{kpc^{-2}\ yr^{-1}}$, respectively, for the four fiducial runs, such that the combinations of (\siggas, \sigSFR) fall along the observed correlation for nearby galaxies at sub-kpc scale \cite[][see Figure 1 of Paper I]{kennicutt98,bigiel08}. The SN rate is related to \sigSFR\ by assuming one SN explodes per 150 $\msun$ of star formation. SNe are placed randomly in the disk, with a scale height of 150 pc for core collapse SNe (90\%), and 325 pc for Type Ia SNe (10\%).  SNe are injected with $10^{51}$ erg thermal energy within a sphere. We use a cooling curve for the solar metallicity gas with a low-temperature cut-off at $T =  300$ K, which assumes collisional ionization equilibrium (CIE) above $10^4$ K \citep{rosen95}. We apply photoelectric heating that scales linearly with the \sigSFR. For the detailed model parameters, see Table 1 of Paper I. We do \textit{not} explicitly include thermal conduction.

To calculate the OVI density and emission, we assume CIE. The relative abundance of O/H is $5\times 10^{-4}$ by number (the ``solar abundance'' in this paper). The fraction of O in the ionization stage of $+5$, as a function of temperature, is taken from \cite{sutherland93}. The collision strength for OVI $\rm{2s\ ^2 S-2p\ ^2 P}$, the transition of which results in the 1032, 1038\AA\ emission, is 5.0 \citep{osterbrock89}. The emission flux of the 1032\AA\ line is twice that of 1038\AA.

\section{Results}

\subsection{OVI-emitting gas in simulations}

\begin{figure}
\begin{center}
\includegraphics[width=0.50\textwidth]{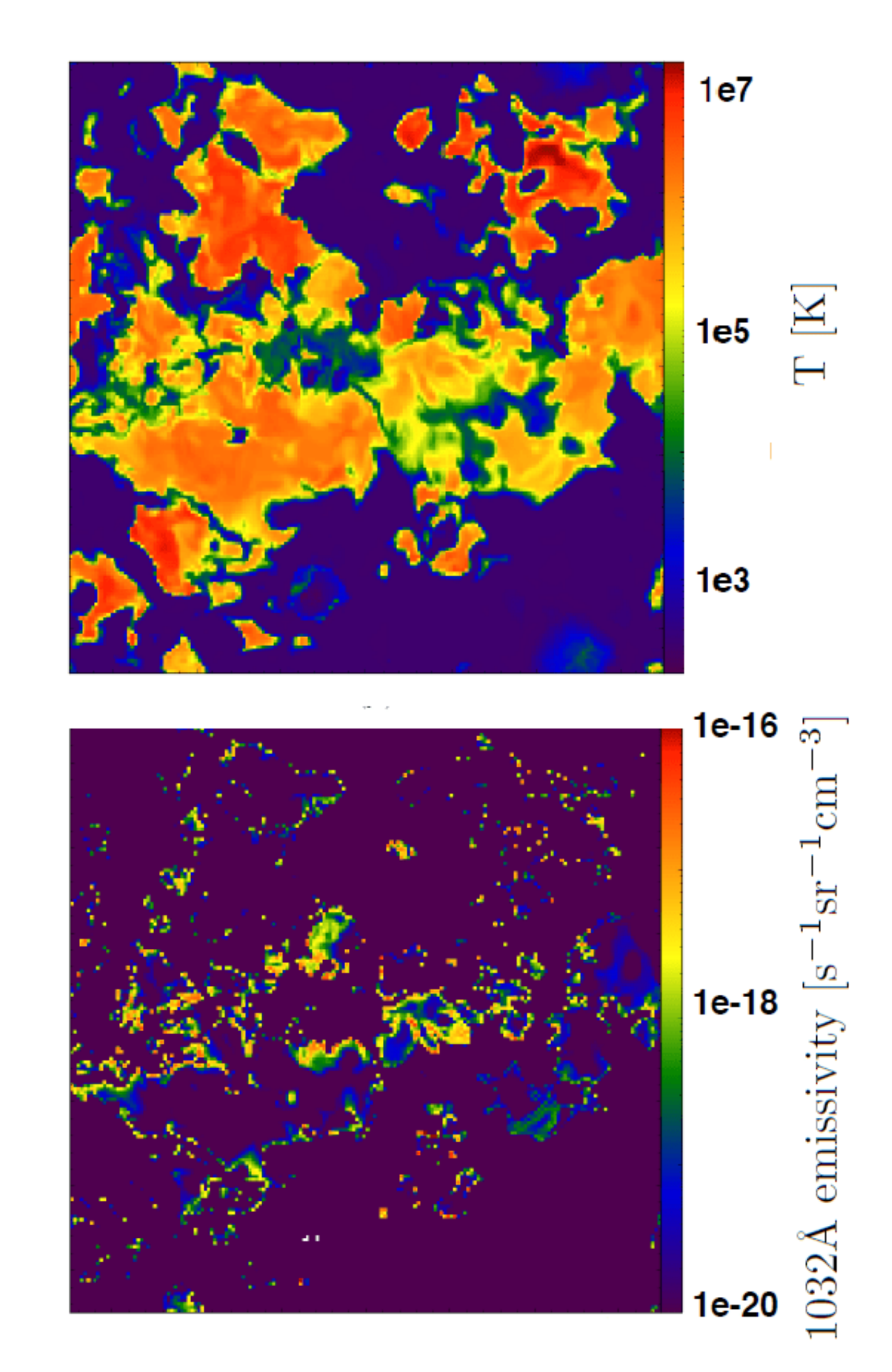}
\caption{Slices of temperature and OVI 1032\AA\  photon emissivity from the midplane of the solar neighborhood model $\Sigma10$-KS. OVI emission mainly comes from the boundary between the hot and cool medium. The OVI-emitting gas is usually not well-resolved, with some emitting regions resolved by one computational cell. The dimensions are 350 pc$\times$ 350 pc, and the resolution is 2 pc.  The color coding for the lower panel has a floor at $10^{-20}\rm{s^{-1} sr^{-1} cm^{-3}} $.   }
\label{f:slice}
\end{center}
\end{figure}

We illustrate the distribution of OVI-emitting gas in simulations in both real and phase space. Figure \ref{f:slice} shows slices of temperature and OVI emissivity from the midplane of the simulation $\Sigma10$-KS. The x-y dimension is 350 pc $\times$ 350 pc, with a resolution of 2 pc. OVI emission mainly originates from the interface between the hot and cool medium. The gas that emits OVI is usually not well-resolved, and some is as thin as a single computational cell. This is generally seen in the simulations, even for the highest-resolution run $\Sigma150$-KS.

\begin{figure}
\begin{center}
\includegraphics[width=0.50\textwidth]{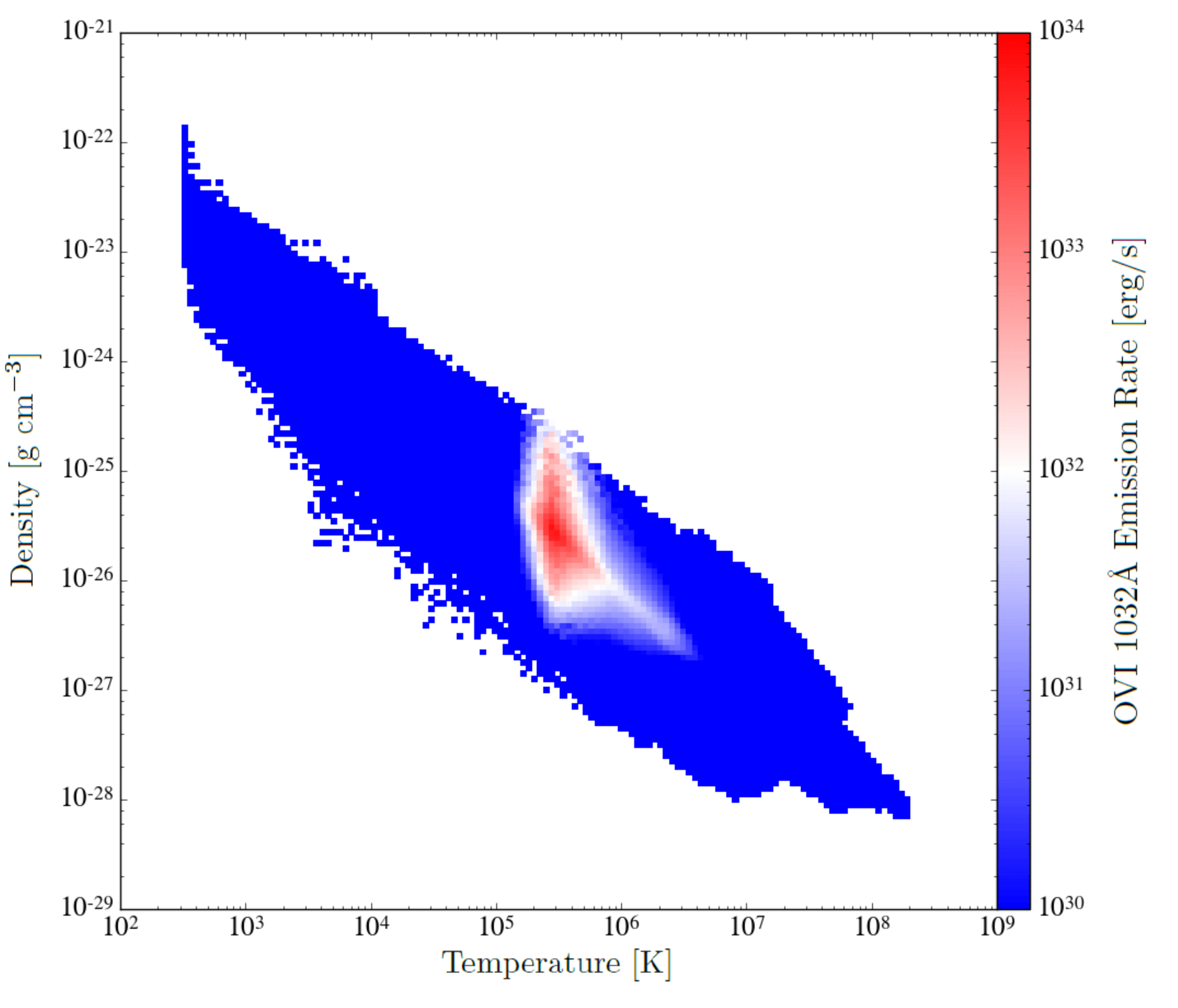}
\caption{Phase diagram for the run $\Sigma$10-KS, where the color coding shows the 1032\AA\ emission rate from gas in that [temperature, density] bin. The calculation includes all gas in the simulation domain. The color coding has a floor at $10^{32}\  \rm{erg\ s^{-1}}$ .}
\label{f:phase_diagram}
\end{center}
\end{figure}

Figure \ref{f:phase_diagram} shows the phase diagram for the same run $\Sigma10$-KS, in which the color coding indicates the total 1032\AA\, emission from each [density, temperature] bin. The majority of the emission is from close to $3\times 10^5$ K, where the ionization fraction of OVI is largest. Gas with intermediate densities,  $0.01-0.1\ \rm{cm^{-3}}$, contributes most of the emission. Below $1.5\times 10^5$ K and above $10^6$ K, the emission is negligible.

\subsection{OVI in the solar neighborhood}

\begin{figure}
\begin{center}
\includegraphics[width=0.50\textwidth]{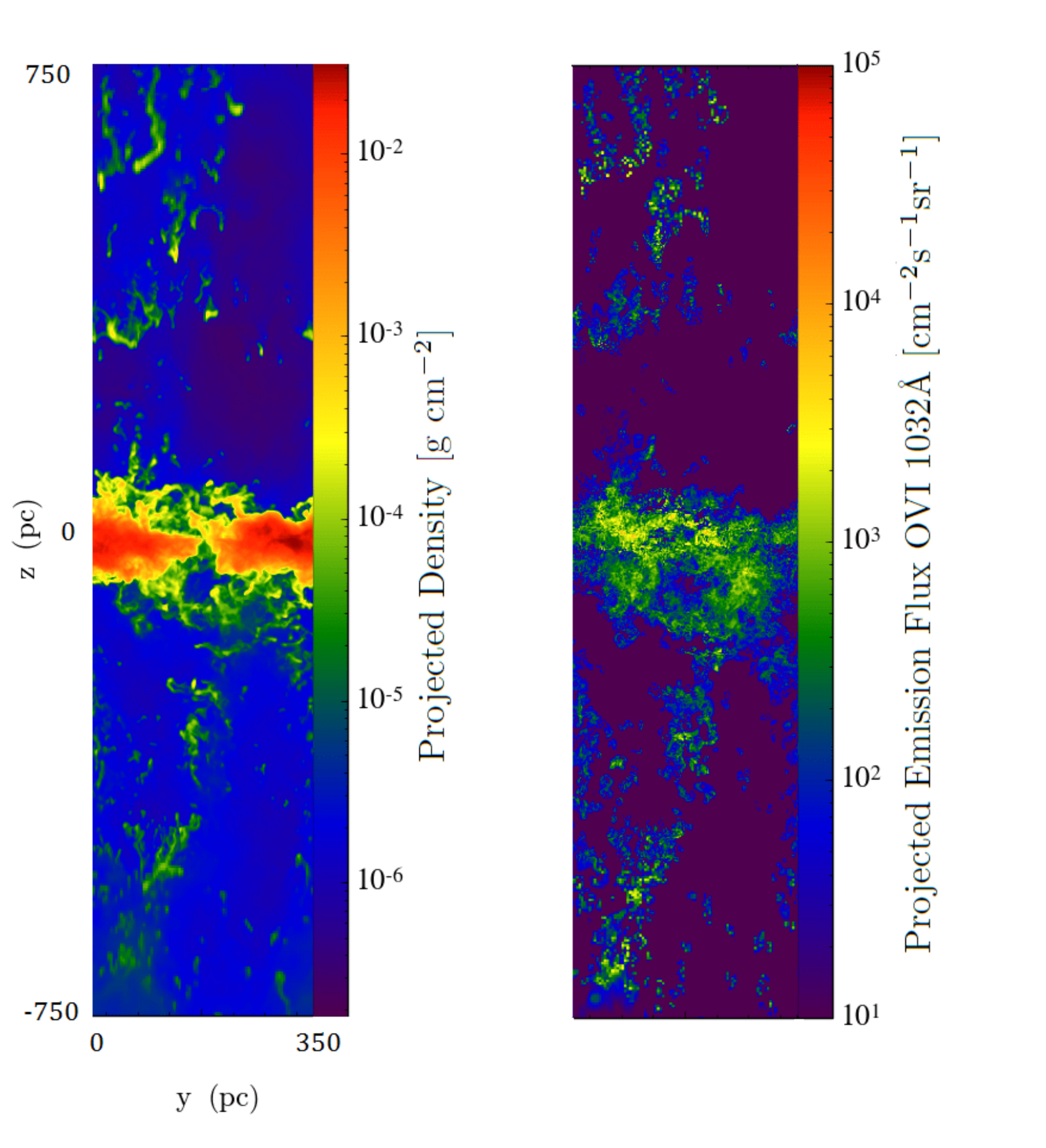}
\includegraphics[width=0.42\textwidth]{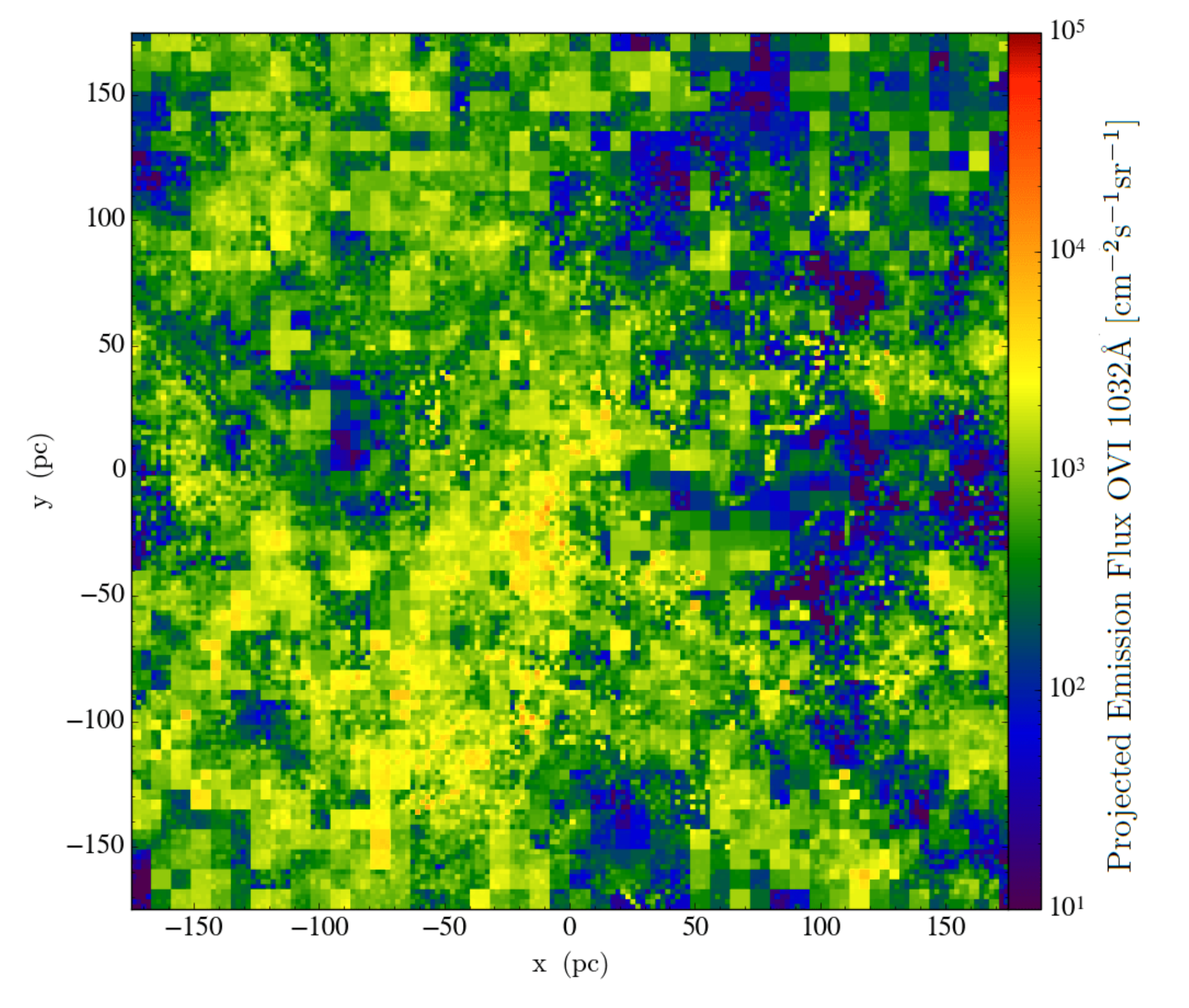}
\caption{Upper left: projected gas density in the x-direction; upper right: projected OVI 1032\AA\ emission in the x-direction (only --0.75 $\leqslant z\leqslant$ 0.75 kpc is shown). Lower: projected OVI 1032\AA\ emission in the z-direction. }
\label{f:projection}
\end{center}
\end{figure}

Figure \ref{f:projection} shows the integrated flux along the x- and the z-directions for the OVI 1032\AA\ emission. The snapshot is taken at $t = 140$ Myr for the run $\Sigma$10-KS. Most of the emission is from near the midplane, where gas is dense and where most SNe explode. Above the plane, gas is mostly hot ($T > 10^6$ K), and the oxygen atoms are ionized even further. It is generally the case that most of the emission is from near the disk in our simulations. Indeed, \cite{hayes16} found emission from the SF disk is much stronger than from the halo. The 1032 \AA\ emission outside of the disk is seen in and around the clouds. The hot gas can shock-compress and ablate the warm clouds, as well as mix with them, all of which may result in gas around $10^{5.5}$ K.  The x-projection of the emission shows some spatial variation. The regions where the emission is weaker have smaller gas column densities, as a result of hot, low-density bubbles that connect to the halo.
 
The \textit{FUSE} satellite has surveyed OVI 1032\AA\ emission from the local ISM at various Galactic latitudes \citep{dixon06, otte06}. The emission is detected for about 40\% of sight lines, with intensities of $(1.9-8.6)\times 10^3\ \rm{photon\  s^{-1}cm^{-2}sr^{-1}}$. \cite{welsh07} used the \textit{SPEAR} satellite to look at the north Galactic pole region, and detected OVI 1032\AA\ emission from 8 out of 16 survey regions, with intensities of $5\times 10^3 - 2\times 10^4\ \rm{photon\  s^{-1}cm^{-2}sr^{-1}}$. Note that the observed emission may be partly from the CGM, so these values may be an upper limit for the emission from near the disk. Our projection plot along the z-direction shows 1032\AA\ emission of $(0.5-6)\times 10^3\ \rm{photon\  s^{-1}cm^{-2}sr^{-1}}$, broadly consistent with the observations. (The emission in the simulations includes the two sides of the midplane, whereas the observation is from within the plane, so there is a factor of $\sim 2$ difference.)

In addition to emission, we also briefly touch on OVI absorption. Our results show very good agreement with the observations. Absorption-line studies from stars with known distances reveal that the local OVI density $n_{\rm{OVI}}$ in the disk to be about $ (2.2\pm 1.0) \times 10^{-8} \rm{cm}^{-3}$ \citep{ jenkins78b, oegerle05, savage06,bowen08, barstow10}. The $n_{\rm{OVI}}$ in our solar neighborhood run is $(2.4\pm 0.3) \times 10^{-8} $ and $(1.8\pm 0.3) \times 10^{-8}\ \rm{cm}^{-3}$, averaged over $|z|\leqslant$ 100 and 200 pc, respectively.   

The non-equilibrium effects will affect these results, generally by boosting the amount of OVI at lower temperatures. Recent simulations by \cite{deavillez12}, which track the ionization stage of oxygen, i.e., relaxing the assumption of CIE, find that up to 70\% (by mass) of the OVI may be at $T\lesssim 10^{5}$ K, although their mean $n_{\rm{OVI}}$ is consistent with ours. We expect that the non-equilibrium effects will affect the OVI absorption more strongly than emission, since low temperatures would lead to less efficient collision and excitation.

\subsection{Scaling of OVI with SFR}
\label{sec:scale}

\begin{figure}
\begin{center}
\includegraphics[width=0.50\textwidth]{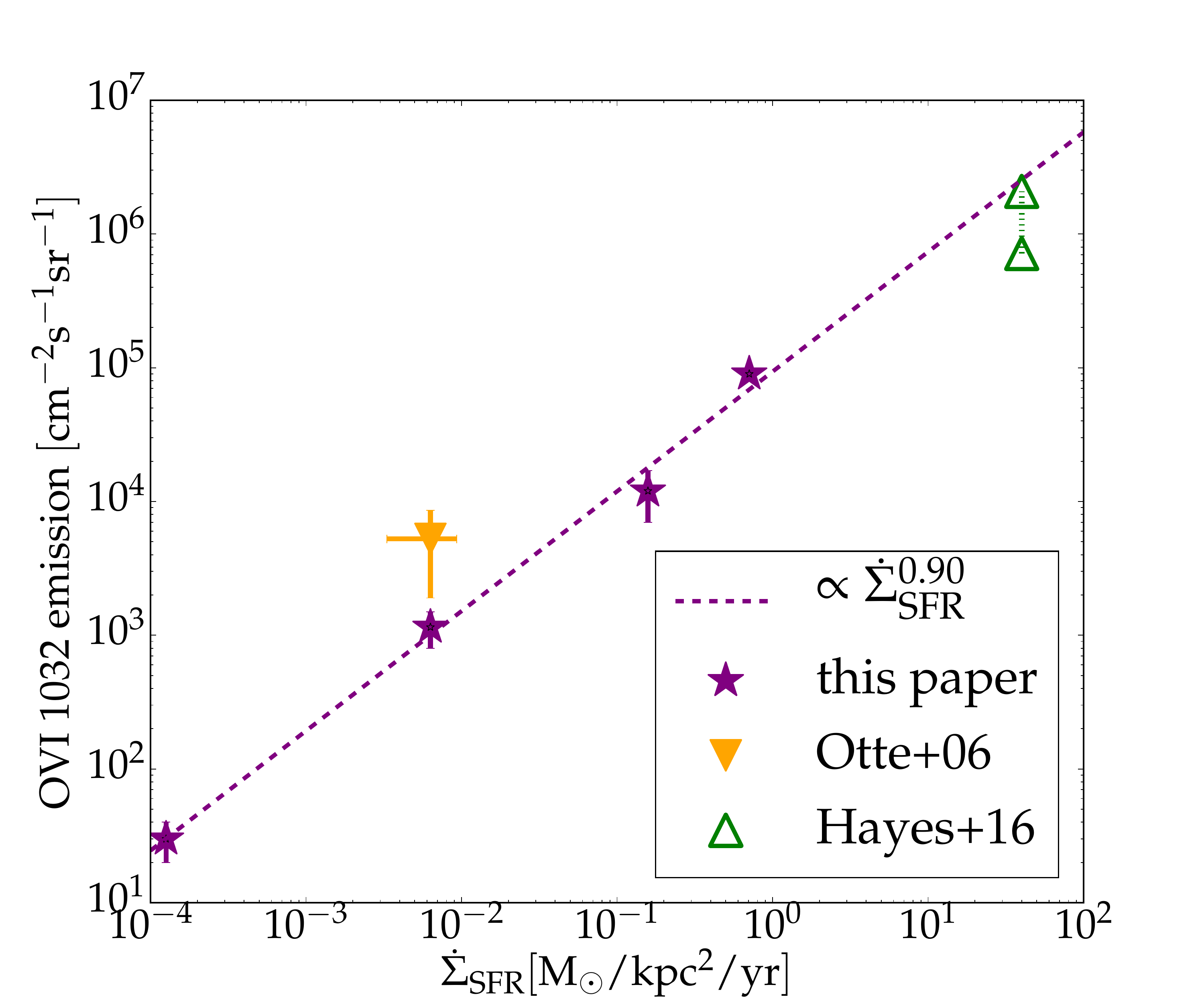}
\caption{Purple stars: OVI 1032\AA\, emission integrated over the z-direction in our simulations, as a function of \sigSFR. The correlation is roughly linear. Filled triangle: observation for the solar neighborhood \citep{otte06}, which detected OVI emission for about 40\% of the sight lines. Hollow triangle: observation of SDSS J1156+5008 \citep{hayes16}, scaled to the solar abundance (the two triangles indicate O/H determined from two methods). }
\label{f:em_OVI}
\end{center}
\end{figure}

Figure \ref{f:em_OVI} shows the OVI 1032\AA\, emission, integrated over the z-direction, as a function of \sigSFR. The quantity is averaged over the x-y plane. The error bar shows the time variation. We find the OVI emission scales roughly linearly with \sigSFR. As mentioned before, the solar neighborhood model agrees well with the observations. 

Observations of OVI emission from external galaxies are limited to a few cases. We show the emission from  SDSS J1156+5008 \citep{hayes16}. The surface brightness reported in that paper is for the 1038\AA\  line. To get that of the 1032\AA\ , we have applied (i) a factor of 2 increase due to the relative intensity of 1032\AA\  and 1038\AA\,; (ii) a factor of 2 increase to account for gas from both sides of the disk, since only the redshifted part is observed as emission. We have also scaled the emission to the solar abundance. \cite{hayes16} calculated O/H using two methods, which differ by a factor of $\sim$3. The two hollow triangles in the plot correspond to these two O/H values. Still, the observed surface brightness is likely a lower limit of the real emission for several reasons. First, the aperture of the COS spectrograph is 2.5 arcsec in diameter, corresponding to a physical scale of r$\sim$9.4 kpc at the galaxy's redshift (0.235). The surface brightness of the 1038\AA\ line is assumed to be uniform inside the aperture, but the star forming region is very compact -- only about 1 kpc across (as seen from H$\alpha$). Thus the actual surface brightness in this starbursting region is probably higher. Second, the 1038\AA\  line can be partially blended with the nearby CII absorption.
Additionally, attenuation by dust in the host galaxy, especially in the disk, can also lower the emission. Based on these arguments, the reported emission intensity is probably a lower limit. Also,  the CGM probably does not contribute much to the surface brightness in this measurement, since for J1156, the OVI emission in the halo is observed, with an intensity much smaller than the center. 

Apart from J1156, another galaxy, Haro 11, turns out to be very similar \citep{grimes07, hayes16}. \cite{hayes16} also compare J1156 with the stacked spectra of other starbursting galaxies from archival data, and find an OVI emission comparable to J1156.

For high \sigSFR, we cannot yet draw a firm conclusion about whether our results are consistent with the (very limited) observations. There can be agreement if, for the very high \sigSFR\ cases, the actual emission is above the observed value for the reasons discussed above, and/or the OVI emission from the simulations increases more slowly for higher \sigSFR, which we do not cover in our simulations.

Finally, the roughly linear scaling of the OVI surface brightness with \sigSFR\  implies that the cooling rate from OVI is an approximately constant fraction of the SNe heating. We find this fraction to be around 0.5\%, including both 1032\AA\ and 1038\AA\ . Therefore cooling from OVI is not an important source of energy loss. Most ($>$ 80\%) cooling of the ISM comes from a broad warm-hot regime: $10^4 -  10^{5.7}$ K. But OVI mainly exists within a narrow fraction of it: $10^{5.3}-10^{5.6}$ K; gas in this temperature range only contributes to a few percent of the total cooling rate.

\begin{figure}
\begin{center}
\includegraphics[width=0.50\textwidth]{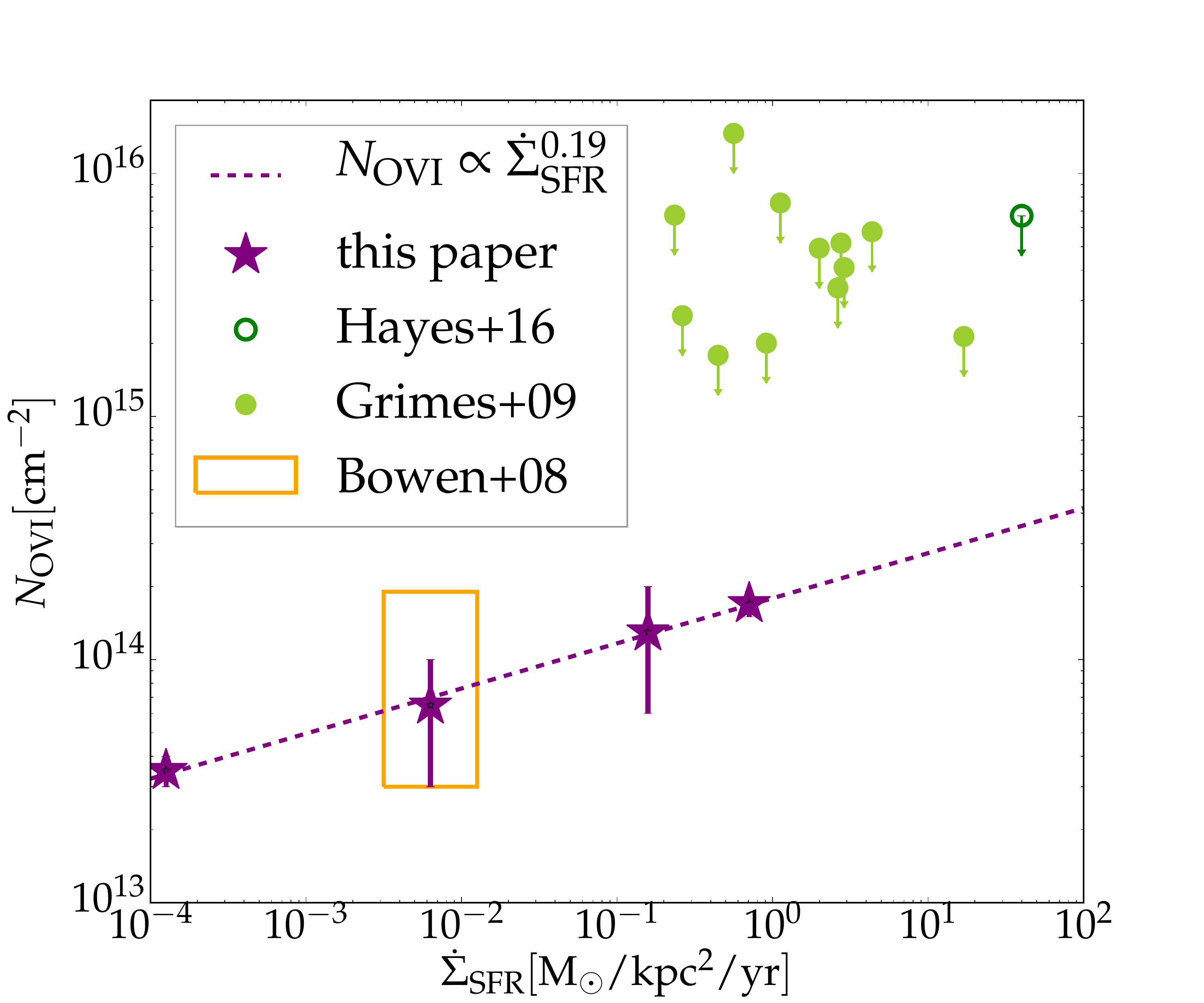}
\caption{Purple stars: mean column density of OVI along the z-direction for the whole simulation domain, as a function of \sigSFR. Orange box: observed value for the solar neighborhood \citep{bowen08}. Filled circles: observed \NOVI for starburst galaxies in \cite{grimes09}. Hollow circle: observed \NOVI for J1156 \citep{hayes16}. For external galaxies, \NOVI have been scaled to the solar abundance; the down arrows indicate these values should be an upper limit for \NOVI near the disk, since they also include OVI in the CGM. }
\label{f:NOVI} 
\end{center}
\end{figure}

The predicted OVI column density \NOVI (across the whole z-direction), as a function of \sigSFR\ , is shown in Figure \ref{f:NOVI} (purple stars). The column density is averaged over the x-y plane. The error bar indicates the time variation. The correlation of \NOVI with \sigSFR\ is positive but the dependence is weak, with a power-law index of 0.19. For four orders of magnitude span in \sigSFR, \NOVI only differs by a factor of 6. 
For the solar neighborhood, we show the observed value as in \cite{bowen08} (orange box), with a factor of 2 increase applied to include OVI from both sides of the galaxy disk; the vertical range of the box indicates the statistical variation of \NOVI, and the horizontal range denotes the uncertainty of \sigSFR. Our result generally agrees with the observation.

\cite{grimes09} have reported \NOVI for 12 starburst galaxies (from absorption against galaxy disks). We plot these \NOVI together with the corresponding \sigSFR\ \citep[Table 1 of ][]{heckman15}. We applied a factor of 2 increase to the data to account for OVI from both sides of the disk, and scale \NOVI to the solar abundance. Additionally, we plot \NOVI for J1156 \citep{hayes16}. Those starburst systems have \NOVI that are usually an order of magnitude larger than (the extrapolation of) our results. But note that a significant fraction of \NOVI can come from the CGM. For nearby Milky Way-like galaxies, \NOVI of the CGM (from absorption against background quasars) ranges from $10^{14}-10^{15} \rm{cm^{-2}}$ \citep{werk16}. The starburst systems may have even higher \NOVI in the CGM for their much stronger SF/outflows activities. If OVI in the CGM indeed dominates the column density, then the extrapolation of our results may be consistent with observations. 

We tested the effect of resolution for the two low-\siggas cases. The test runs include a midplane resolution of 4 pc for $\Sigma$10-KS (fiducial 2 pc), and 2.5 pc and 10 pc for $\Sigma$1-KS (fiducial 5 pc). The 1032\AA\ emission agrees within the time variation (which can be as large as 80\%). The mean values of OVI emission increase with improved resolution.  Such a trend can be partly attributed to that resolving more cloud fractals increases the total area of boundary layers. The convergence rate is faster for $\Sigma$1-KS than $\Sigma$10-KS: improving resolution by a factor of 2 leads to 20\% and 60\% increase of the mean OVI emission, respectively. The total cooling rate shows less time variation ($\sim$20\%), and is consistent for different resolutions as well.

\section{Conclusions}

In this Letter we calculate the OVI 1032\AA\ emission from our ISM simulations with SN feedback. This is done to test the radiative cooling from gas with $T\sim 10^{5.5}$ K (which cools very efficiently but is usually not very well-resolved in simulations) against observations. We find that, for the solar neighborhood, our results agree well with the observations, both in terms of emission flux and mean OVI density in absorption.  Most emission comes from the gas disk. The relatively dense (0.01-0.1 $\rm{cm^{-3}}$) gas around the interface between hot gas and warm clouds contributes the majority of the emission. 

Changing \siggas and \sigSFR\ along the Kennicutt relation, we find that both \NOVI and the OVI surface brightness increase with \sigSFR. For \NOVI, the dependence is quite weak, while the surface brightness of OVI emission scales roughly linearly with \sigSFR. OVI emission is approximately 0.5\% of the SNe heating rate. It cannot yet be determined definitely whether our results give a reasonable emission for high \sigSFR, because of the limited observational and numerical samples, although agreement is not unlikely. More observations of the OVI emission, for both low and intermediate-high \sigSFR, are needed for a more complete comparison.

\section*{Acknowledgement}
We thank the referee for constructive comments. We thank Ed Jenkins, Jules Halpern, Matthew Hayes and David Schiminovich for reading the manuscript and giving invaluable feedback. We thank Eve Ostriker for very helpful discussions. Computations were performed using the publicly-available Enzo code. Data analysis and visualization are partly done using \textsf{yt} \citep{turk11}. We acknowledge NASA grant NNX15AB20G , NSF grants AST-1312888 and AST-1615955. This work used the XSEDE Stampede cluster (NSF grant ACI-1053575), NASA HEC Pleiades cluster, and Columbia University's Yeti Cluster. \\ 

\vspace{0.2in}


\end{document}